\documentclass[aps,showpacs,amsmath,amssymb,twocolumn,pra,superscriptaddress,notitlepage]{revtex4-2}

\usepackage{physics}
\usepackage{mathtools}

\DeclarePairedDelimiter{\ceil}{\lceil}{\rceil}
\newcommand{\mr}[1]{\mathrm{#1}}

\usepackage{xcolor}
\usepackage[colorlinks=true, allcolors=blue]{hyperref}

\begin{document} 

\title{Tensor-based phase difference estimation on time series analysis}

\author{Shu Kanno}
\email{shu.kanno@quantum.keio.ac.jp}
\affiliation{Mitsubishi Chemical Corporation, Science \& Innovation Center, Yokohama, 227-8502, Japan}
\affiliation{Quantum Computing Center, Keio University, 3-14-1 Hiyoshi, Kohoku-ku, Yokohama, 223-8522, Japan}

\author{Kenji Sugisaki}
\affiliation{Deloitte Tohmatsu LLC, 3-2-3 Marunouchi, Chiyoda-ku, Tokyo 100-8363, Japan}
\affiliation{Quantum Computing Center, Keio University, 3-14-1 Hiyoshi, Kohoku-ku, Yokohama, 223-8522, Japan}
\affiliation{Centre for Quantum Engineering, Research and Education, TCG Centres for Research and Education in Science and Technology, Sector V, Salt Lake, Kolkata 700091, India}

\author{Rei Sakuma}
\affiliation{Materials Informatics Initiative, RD Technology \& Digital Transformation Center, JSR Corporation, 3-103-9 Tonomachi, Kawasaki-ku, Kawasaki, 210-0821, Japan}
\affiliation{Quantum Computing Center, Keio University, 3-14-1 Hiyoshi, Kohoku-ku, Yokohama, 223-8522, Japan}

\author{Jumpei Kato}
\affiliation{Mitsubishi UFJ Financial Group, Inc. and MUFG Bank, Ltd., 4-10-2 Nakano, Nakano-ku, Tokyo 164-0001, Japan}
\affiliation{Quantum Computing Center, Keio University, 3-14-1 Hiyoshi, Kohoku-ku, Yokohama, 223-8522, Japan}
\affiliation{Graduate School of Science and Technology, Keio University, 3-14-1 Hiyoshi, Kohoku, Yokohama, Kanagawa 223-8522, Japan}

\author{Hajime Nakamura}
\affiliation{Quantum Computing Center, Keio University, 3-14-1 Hiyoshi, Kohoku-ku, Yokohama, 223-8522, Japan}

\author{Naoki Yamamoto}
\affiliation{Quantum Computing Center, Keio University, 3-14-1 Hiyoshi, Kohoku-ku, Yokohama, 223-8522, Japan}
\affiliation{Department of Applied Physics and Physico-Informatics, Keio University, 3-14-1 Hiyoshi, Kohoku-ku, Yokohama 223-8522, Japan}

\begin{abstract}
We propose a phase-difference estimation algorithm based on the tensor-network circuit compression, leveraging time-evolution data to pursue scalability and higher accuracy on a quantum phase estimation (QPE)-type algorithm. Using tensor networks, we construct circuits composed solely of nearest-neighbor gates and extract time-evolution data by four-type circuit measurements. In addition, to enhance the accuracy of time-evolution and state-preparation circuits, we propose techniques based on algorithmic error mitigation and on iterative circuit optimization combined with merging into matrix product states, respectively.
Verifications using a noiseless simulator for the 8-qubit one-dimensional Hubbard model using an ancilla qubit show that the proposed algorithm achieves accuracies with 0.4--4.7\% error from a true energy gap on an appropriate time-step size, and that accuracy improvements due to the algorithmic error mitigation are observed. We also confirm the enhancement of the overlap with matrix product states through iterative optimization. Finally, the proposed algorithm is demonstrated on IBM Heron devices with Q-CTRL error suppression for 8-, 36-, and 52-qubit models using more than 4,000 2-qubit gates. These largest-scale demonstrations for the QPE-type algorithm represent significant progress not only toward practical applications of near-term quantum computing but also toward preparation for the era of error-corrected quantum devices.
\end{abstract}

\maketitle

\section{Introduction \label{sec:introduction}}
Quantum computers can perform chemical tasks that are intractable for classical algorithms. 
One of the tasks is calculating spectral quantities, such as energies and energy gaps, in chemical and physical models. The algorithms for the tasks include a quantum phase estimation (QPE)-type algorithms, which estimate the value $\Theta$ from the phase of $\exp(i\Theta)$ constructed in a quantum circuit~\cite{Yu_Kitaev1995-cf, Cleve1998-cg, Nielsen2010-hw}: if $\Theta$ corresponds to an eigenvalue for a given Hamiltonian, the algorithm estimates energy.
Since the QPE-type algorithms can estimate the quantities with desired precision with a polynomial cost for a system size, a clear exponential advantage is expected by a quantum computer in theory; in the classical counterpart, the full configuration interaction (FCI), the computational cost increases exponentially with the system size, and benchmarks have been performed only up to 46 qubits at maximum~\cite{Gao2024-aw}. 

In the progress history for the QPE-type algorithm~\cite{higgins2007entanglement, Kimmel2015-hy, Wiebe2016-kr}, approaches based on time-series data analysis~\cite{Somma2019-gb, O-Brien2019-xm, Dutkiewicz2022-zd} have attracted attention in the near-term or early fault-tolerant quantum computer (FTQC) era.
Specifically, in this approach, the phase $\Theta$ is estimated from time series data $\{\exp(i\Theta t) \}$ for multiple times $t$. 
The approach leverages mature classical processing techniques such as the Fourier transformation, the convolution integral~\cite{Lin2022-rn, Wang2022-yo}, and direct data fitting~\cite{Ding2023-iu, Ding2023-aa, Ding2023-rs}, leading to shallow circuits.
However, on current noisy devices, experimental implementations have been limited to only a few qubit systems~\cite{Blunt2023-jf} since the circuit to run the approach is still too deep for current noisy devices.
For the end-to-end implementation on the devices, the efficient implementation of initial state preparation, time evolution operators, and time series data extraction is required.

\begin{figure*}[]
 \centering
 \includegraphics[width=1\textwidth]{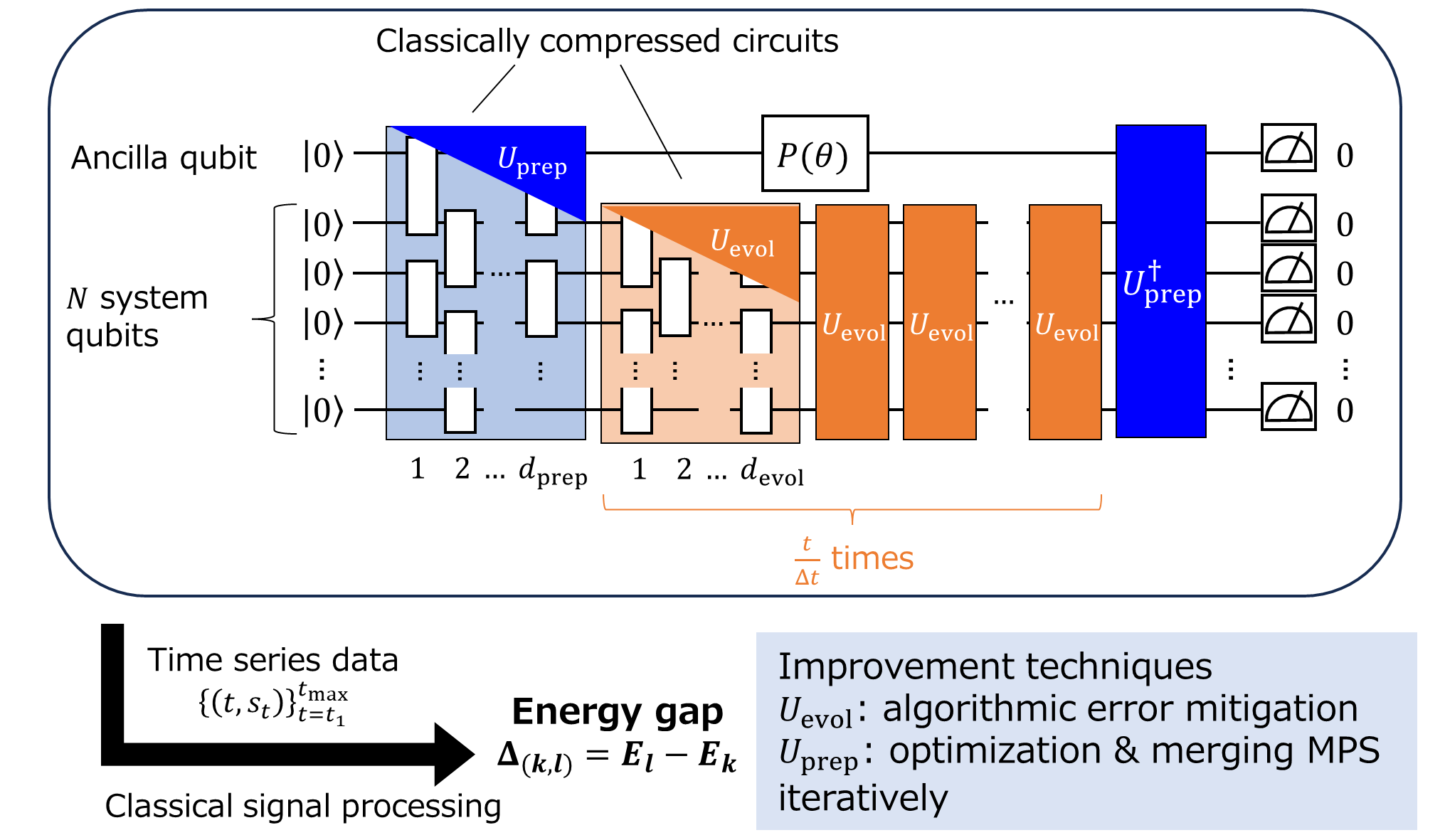}
\caption{Overview of our proposal. $P(\theta) =
\begin{pmatrix}
1 & 0 \\
0 & e^{i\theta}
\end{pmatrix}$ represents the phase gate with the parameter $\theta$ taking $0, \pi/2, \pi,$ and $3\pi/2$. $s_t$ is a signal to construct a time series data $\{(t,s_t )\}_{t=t_1}^{t_\mr{max}}$, where $t_1$ and  $t_\mr{max}$ are $\Delta t$ and a maximum time, respectively. } 
 \label{fig: overview_2column.png}
\end{figure*}

We propose an efficient QPE-type algorithm using time-series analysis, as shown in Fig.~\ref{fig: overview_2column.png}.  
Here, we adopt a strategy combining tensor-network circuit compression~\cite{Rudolph2024-jo, Ran2020-li, Lin2021-km, Dborin2022-mh, Dborin2022-sx, Mc_Keever2023-bu, Anselme_Martin2024-dw, Causer2024-wd, Shirakawa2024-yi, Mc_Keever2024-ji, D-Anna2025-wk, Sugawara2025-xu, Gibbs2025-hd, Le2025-ne, Mansuroglu2026-ay, Karacan2026-zh} and quantum phase-difference estimation (QPDE)~\cite{Sugisaki2021-gg}, which enabled our group to achieve the largest demonstration for the QPE-type algorithm up to 32-qubit system~\cite{Kanno2025-tt}: 
QPDE is one QPE-type algorithm focusing on energy gap calculation (i.e., $\Theta$ corresponds to the gap $\Delta$) without controlled operations for time evolution operators. The circuits for initial-state preparation and time evolution in QPDE are further compressed to sets of brick-wall gates through the matrix product states (MPS) and matrix product operators (MPO), represented as $U_\mr{prep}$ and $U_\mr{evol}$, respectively. We choose the direct data fitting as the classical signal processing~\cite{Ding2023-iu}.

In addition, to improve the accuracy of tensor-based QPDE, we introduce two new refinement techniques. The first is the improvement of time evolution error by using the algorithmic error mitigation (AEM)~\cite{Endo2019-oh}. While the AEM is mainly targeted to mitigate Trotter errors, our approach, which uses techniques on Ref.~\cite{Robertson2024-jv}, provides a framework for mitigating approximation errors introduced by tensor compression of a time evolution circuit. 
The second is the improvement of a state-preparation overlap. The state-preparation circuit is optimized to approximate an MPS, but the classical optimization cost grows exponentially with the circuit depth. In this work, we propose an approach that allows the circuit depth to be increased while suppressing this exponential growth in classical cost.

Finally, we executed time-series QPDE on real quantum hardware for a one-dimensional Hubbard model with up to 52 qubits, exceeding the size limit of FCI.
Although the accuracy in this approach would be confined depending on the tensor-network topology, e.g., quasi-one dimension in the MPS and MPO, demonstrations for the large-scale QPE-type algorithm on current devices create a clear pathway for the discovery not only for near-term quantum-advantage tasks but also for practical insights toward the FTQC era.
This point is also clearly distinct from recent large-scale demonstrations of energy estimation based on quantum–classical hybrid approaches~\cite{Kanno2023-ip, Robledo-Moreno2025-ul}.
Note that in our previous demonstration~\cite{Kanno2025-tt}, we adopted an adaptive approach based on Bayesian inference. In addition to the advantages described above, the present non-adaptive time-series approach has the benefit that data can be acquired at all time steps until the signal decays, and that the number of shots required per time step is small.

\begin{figure}[]
 \centering
 \includegraphics[width=1\columnwidth]{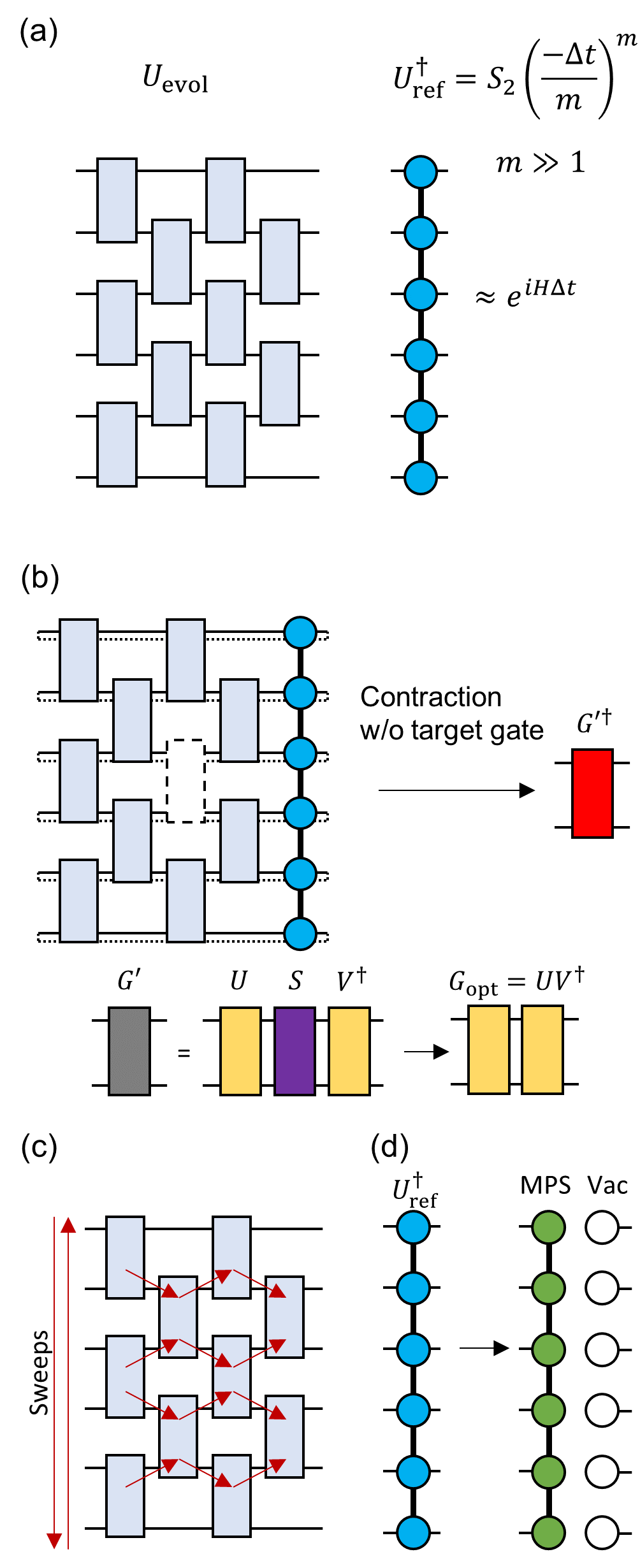}
\caption{Summary of the unitary compression in our previous work. (a) Initial preparations for $U_\mr{evol}$ and $U_\mr{ref}^{\dagger}$. (b) Gate optimization for a target gate. (c) Optimization sequence for gates. (d) Substitution for $U_\mr{prep}$.} 
 \label{fig: UnitaryCompression_1column.png}
\end{figure}

\section{Method}
As in Fig.~\ref{fig: overview_2column.png}, we extract a signal $s_t$, having energy gap information, for a time $t$ from the compressed circuit, and then calculate the energy gap from the data using classical signal processing, where the time series data is denoted as $\{(t,s_t )\}_{t=t_1}^{t_\mr{max}}$, and $t_1$ and  $t_\mr{max}$ are $\Delta t$ and a maximum time, respectively.
The circuit applies the state preparation operator as $U_{\mathrm{prep}}\ket{0}^{\otimes N+1}$, a phase gate $P(\theta)$ having a phase $\theta$, the time evolution operator $U_\mathrm{evol}$, and the inverse operator $U_{\mathrm{prep}}^\dagger$, and finally measures by all zeros.
$U_\mathrm{prep}$ and $U_\mathrm{evol}$ are composed of the brick-wall gates with depths $d_\mathrm{prep}$ and $d_\mathrm{evol}$, respectively.

We first explain the extraction of $s_t$ in the next section. Then we show a summary of a procedure for compression of $U_\mr{prep}$ and $U_\mr{evol}$ proposed by the previous study~\cite{Kanno2025-tt}. We also describe the techniques to improve the quality for $U_\mr{prep}$ and $U_\mr{evol}$.

\subsection{Time series data calculation}
We generally describe the prepared state as
\begin{equation}
\begin{aligned}
    U_{\mathrm{prep}}\ket{0}^{\otimes N+1}
    &= a_0 \ket{0}\ket{\phi} + a_1 \ket{1}\ket{\phi'}, \\
\end{aligned}
\end{equation}
where $a_0, a_1 \in \mathbb{C}$, $|a_0|^2 + |a_1|^2 = 1$, and $\ket{\phi}$ and $\ket{\phi'}$ are states for a system.
The expectation values of the circuits as a function of \(\theta\) are
\begin{equation}
\begin{aligned}
    m(\theta) &= \Tr[(P(\theta)^\dagger U_{\mathrm{prep}} \ket{0}\bra{0}^{\otimes N+1}  U_{\mathrm{prep}}^\dagger P(\theta))\rho(t)]\\
    &= |a_0|^4 \abs{\mel{\phi}{(U_{\mathrm{evol}})^{\frac{t}{\Delta t}}}{\phi}}^2\\
    &+ e^{i\theta}|a_0|^2|a_1|^2 \mel{\phi'}{(U_{\mathrm{evol}})^{\frac{t}{\Delta t}}}{\phi'} \mel{\phi}{(U_{\mathrm{evol}}^\dagger)^{\frac{t}{\Delta t}}}{\phi} \\
    &+ e^{-i\theta}|a_0|^2|a_1|^2 \mel{\phi}{(U_{\mathrm{evol}})^{\frac{t}{\Delta t}}}{\phi} \mel{\phi'}{(U_{\mathrm{evol}}^\dagger)^{\frac{t}{\Delta t}}}{\phi'}\\
    &+ |a_1|^4 \abs{\mel{\phi'}{(U_{\mathrm{evol}})^{\frac{t}{\Delta t}}}{\phi'}}^2,
    \label{eq: m(theta)}
\end{aligned}
\end{equation}
where $\rho(t) = \ket{\psi(t)}\!\bra{\psi(t)}$, and $\ket{\psi(t)} = (U_{\mathrm{evol}})^{t/\Delta t} U_{\mathrm{prep}} \ket{0}^{\otimes N+1}$.

By combining \(m(\theta)\) for \(\theta = 0, \pi/2, \pi\), and \(3\pi/2\), yielding a time signal data $s_t$ as
\begin{equation}
\begin{aligned}
\label{eq: nondiagonal extraction}
    s_t & =\frac{1}{4|a_0|^2(1 - |a_0|^2)}\\
    &\times\left\{ m(0) - m(\pi)
    - i \left[ m\!\left( \frac{\pi}{2} \right)
    - m\!\left( \frac{3\pi}{2} \right) \right] \right\}\\
    &= \mel{\phi}{(U_{\mathrm{evol}}^\dagger)^{\frac{t}{\Delta t}}}{\phi} \mel{\phi'}{(U_{\mathrm{evol}})^{\frac{t}{\Delta t}}}{\phi'},
\end{aligned}
\end{equation}
where the first (the last) two terms in the second line correspond to the real (imaginary) part of $s_t$.
\(|a_0|^2\) can be classically determined since we can have $U_{\mathrm{prep}}\ket{0}^{\otimes N+1}$ as an MPS and $|a_0|^2 = |(\bra{0} \otimes I^{\otimes{N}} )U_{\mathrm{prep}}\ket{0}^{\otimes N+1}|^2$.

Let $\ket{\psi_k}$ be an eigenstate such that \(H\ket{\psi_k} = E_k\ket{\psi_k}\) with eigenvalues \(E_0 < E_1 < E_2 < \cdots\), where $\ket{\psi_0} = \ket{\psi_\mr{g}}$ and $\ket{\psi_1} = \ket{\psi_\mr{ex}}$.
Expanding the states as
\begin{equation}
\begin{aligned}
    \ket{\phi} &= \sum_k \sqrt{p_k} e^{i\xi_k} \ket{\psi_k}, \\
    \ket{\phi'} &= \sum_l \sqrt{q_l} e^{i\varphi_l} \ket{\psi_l},
\end{aligned}
\end{equation}
we obtain the spectral form
\begin{equation}
\begin{aligned}
    s_t &= \mel{\phi}{(U_{\mathrm{evol}}^\dagger)^{\frac{t}{\Delta t}}}{\phi} \mel{\phi'}{(U_{\mathrm{evol}})^{\frac{t}{\Delta t}}}{\phi'}\\
    &= \sum_{k,l} p_k q_l e^{-it(E_l - E_k)} \\
    &= \sum_J P_J e^{-i\Delta_J t},
\end{aligned}
\end{equation}
where indices $k$ and $l$ are merged as $J = (k,l)$, \(\Delta_{J} = E_l - E_k\), and \(P_J = p_k q_l \ge 0\) denotes the corresponding spectral weight.

When \(\ket{\phi} \approx \ket{\psi_{\mathrm{g}}}\) (ground state)
and \(\ket{\phi'} \approx \ket{\psi_{\mathrm{ex}}}\) (excited state),
the dominant frequency component with the largest \(P_J\),  \(\Delta_J\),
corresponds to the target energy gap \(\Delta_{J=(0,1)} = E_1 - E_0\), where we choose the first excited state as the target excited state.
Note that we can extract the information for the other gaps by adjusting $\ket{\psi_\mr{g}'}$ and $\ket{\psi_\mr{ex}'}$ in the MPS preparation of Eq.~\eqref{eq: prep}.

We explain the affection for deporizing noise.
The state before the measurement is
\begin{equation}
\begin{aligned}
    \tau(t) = U_{\mathrm{prep}}^\dagger P(\theta)\rho(t) P(\theta)^\dagger U_{\mathrm{prep}}.
\end{aligned}
\end{equation}

Due to a depolarizing noise channel, the state transforms as
\begin{equation}
\begin{aligned}
    \tau'(t) = (1 - p_{\mathrm{dep}}(t))\tau(t) + \frac{p_{\mathrm{dep}}}{2^{N+1}} I^{\otimes N+1} ,
\end{aligned}
\end{equation}
where \( p_{\mathrm{dep}} \) is an error rate.

The measured probability \( m'(\theta) \) is then
\begin{equation}
\begin{aligned}
    m'(\theta)
    &= \Tr\left[ \ket{0}\bra{0}^{\otimes N+1} \tau'(t) \right] \\
    &= (1 - p_{\mathrm{dep}}(t)) \Tr\left[ \ket{0}\bra{0}^{\otimes N+1} \tau(t) \right]\\ 
    &+ \frac{p_{\mathrm{dep}}}{2^{N+1}} \Tr\left[ \ket{0}\bra{0}^{\otimes N+1} I^{\otimes N+1} \right]\\
    &= (1 - p_{\mathrm{dep}}(t)) m(\theta)
    + \frac{p_{\mathrm{dep}}(t)}{2^{N+1}} .
    \label{eq:m'(theta)}
\end{aligned}
\end{equation}

By combining \(m(\theta)\) at \(\theta = 0, \pi/2, \pi, 3\pi/2\),
\begin{equation}
\begin{aligned}
    s_t = (1 - p_{\mathrm{dep}}(t)) \mel{\phi}{(U_{\mathrm{evol}}^\dagger)^{\frac{t}{\Delta t}}}{\phi} \mel{\phi'}{(U_{\mathrm{evol}})^{\frac{t}{\Delta t}}}{\phi'},
\end{aligned}
\end{equation}
which is the same as Eq.~\eqref{eq: nondiagonal extraction} except for the amplitude decay \((1 - p_{\mathrm{dep}}(t))\).

The gap $\Delta_J$ can be extracted from the time series data $s_t$ by using classical processing. We adopt the data fitting~\cite{Ding2023-iu} with the matrix pencil method~\cite{Sarkar1980-nc, Steffens2017-zz, Dutkiewicz2022-zd} as an initial guess for the classical signal processing. See Appendix~\ref{sec: eigenvalue extraction from time series data} for details.

\subsection{Summary of classical unitary compression using tensor networks\label{sec: summary of classical unitary compression using tensor networks}}
The time-evolution operator $U_{\mathrm{evol}}$ and the state-preparation circuit $U_{\mathrm{prep}}$ are optimized in a brick-wall ansatz so that they satisfy
\begin{equation}
\begin{aligned}
U_{\mathrm{evol}} &\approx e^{-i H \Delta t},
\end{aligned}
\end{equation}
and
\begin{equation}
\begin{aligned}
U_{\mathrm{prep}} \ket{0}^{\otimes N+1}
&\approx 
\frac{1}{\sqrt{2}}
\left(
    \ket{0}\ket{\psi_{\mathrm{g}}}
    +
    \ket{1}\ket{\psi_{\mathrm{ex}}}
\right),
\end{aligned}
\end{equation}
where $\Delta t$ is the time step, $N$ is the number of system qubits, and $\ket{\psi_{\mathrm{g}}}$ and $\ket{\psi_{\mathrm{ex}}}$ denote the ground and excited states, respectively.

We first describe the treatment of time evolution $U_\mathrm{evol}$ in Fig.~\ref{fig: UnitaryCompression_1column.png} (a)--(c). As a reference, we prepare an accurate MPO representation of the time-evolution operator, denoted $U_{\mathrm{ref}}^\dagger$  in Fig.~\ref{fig: UnitaryCompression_1column.png}(a). Assuming $H = \sum_{\beta=1}^K C_{\beta} P_{\beta}$, $C_{\beta}$ is a coefficient, and $P_{\beta}$ is a Pauli term, we construct $U_{\mathrm{ref}}^\dagger$ as
\begin{equation}
\begin{aligned}
    U_{\mathrm{ref}}^\dagger &= S_2\!\left( \frac{-\Delta t}{m} \right)^m\\
    &=\left\{\prod_{\beta = 1}^{K} \exp (iC_{\beta} P_{\beta} \frac{\Delta t}{2m}) \prod_{\beta = K}^{1} \exp (iC_{\beta} P_{\beta}\frac{\Delta t}{2m})\right\}^m,
    \label{eq: reference time evolution}
\end{aligned}    
\end{equation}
where $S_2(\cdot)$ is the second-order Trotter formula, and $m$ is the number of time slices. 

We then optimize each of the two-qubit gates in the brick-wall circuit $U_\mathrm{evol}$ to minimize 
\begin{equation}
    \begin{aligned}
        \norm{U_{\mathrm{ref}}^\dagger - U_{\mathrm{evol}}}_\mr{F}^2 &= \Tr[U_{\mathrm{ref}}^{\dagger}U_{\mathrm{ref}}] + \Tr[U_{\mathrm{evol}}^{\dagger}U_{\mathrm{evol}}]\\
        &- 2\Re\Tr[U_{\mathrm{ref}}^{\dagger}U_{\mathrm{evol}}]\\
        &= 2^{N+1}-2\Re\Tr[U_{\mathrm{ref}}^{\dagger}U_{\mathrm{evol}}],
    \end{aligned}
\end{equation}
where $\norm{\cdot}_\mr{F}$ is the Frobenius norm.
Specifically, we first contract all of the tensors except for the target gate and obtain a four-leg tensor $G'^{\dagger}$ in Fig.~\ref{fig: UnitaryCompression_1column.png}(b). After the singular value decomposition (SVD) for $G'$ as $G'=USV^{\dagger}$, we obtain the local optimal gate $G_{\mathrm{opt}}=UV^{\dagger}$ by substituting $S$ to identity, where $U$ and $V^{\dagger}$ are unitary operators.
This optimization is performed for each gate, where the gate optimization is performed in the order of sweeping up and down the zigzag path from left to right in Fig.~\ref{fig: UnitaryCompression_1column.png}(c). 
The initial gate was constructed by perturbing the identity operator.

The procedure for the state preparation is almost the same as that for the time evolution; we simply substitute an $N$ qubit brick-wall circuit $U_\mathrm{evol}$ to an $N+1$ qubit brick-wall circuit $U_\mathrm{prep}$, and $U_\mathrm{ref}^\dagger$ to $\ket{0}^{\otimes N+1} \bra{\mathrm{MPS}}$ in Fig.~\ref{fig: UnitaryCompression_1column.png}(d). 
Here, 
\begin{equation}
    \begin{aligned}
    \ket{\mathrm{MPS}} = \frac{1}{\sqrt{2}}(\ket{0}\ket{\psi_{\mathrm{g}}'} + \ket{1}\ket{\psi_{\mathrm{ex}}'}),
    \label{eq: prep}
    \end{aligned}
\end{equation}
which approximates $\frac{1}{\sqrt{2}}(\ket{0}\ket{\psi_{\mathrm{g}}} + \ket{1}\ket{\psi_{\mathrm{ex}}})$, where the approximate ground and excited states, $\ket{\psi_{\mathrm{g}}'}$ and $\ket{\psi_{\mathrm{ex}}'}$, respectively, are prepared by the density matrix renormalization group (DMRG)~\cite{White1992-dk, Schollwock2011-im} as MPSs. 
The superposition state in Eq.~\eqref{eq: prep} can be constructed from the MPSs, $\ket{\psi_{\mathrm{g}}'}$ and $\ket{\psi_{\mathrm{ex}}'}$~\cite{Kanno2025-tt}, so we obtain $\ket{\mr{MPS}}$ through the normalization (and left orthogonalization) of $\ket{0}\ket{\psi_{\mathrm{g}}'} + \ket{1}\ket{\psi_{\mathrm{ex}}'}$.

Finally, we note that if $\ket{{\psi_{\mathrm{ex}}'}}$ is replaced by $\ket{0}^{\otimes N}$, an absolute ground state energy value can be calculated rather than the gap~\cite{Sugisaki2021-nk, Kanno2025-tt}. However, we checked numerically that the accuracy in the optimization of $U_\mr{prep}$ was poor, particularly for large-scale systems, and therefore, we do not investigate it in the present study.

\subsection{Algorithmic error mitigation for a compressed time evolution circuit}
We construct the AEM formalism for a compressed time evolution circuit based on the dynamical multi-product formula~\cite{Robertson2024-jv}.  
Let \(U_\mathrm{evol}^i \) be an operator approximating the time evolution of $\Delta t$ with index $i$.
The precise time-evolved density matrix $\rho(t)$ is expressed as
\begin{equation}
\begin{aligned}
    \rho_{\mathrm{p}}(t) = \ket{\psi_{\mathrm{p}}(t)}\!\bra{\psi_{\mathrm{p}}(t)}, 
\end{aligned}
\end{equation}
and
\begin{equation}
\begin{aligned}
    \ket{\psi_{\mathrm{p}}(t)} = e^{-iHt} \ket{\psi_{\mathrm{prep}}},
\end{aligned}
\end{equation}
where $e^{-iHt}$ is a precise time evolution operator, $\ket{\psi_{\mathrm{prep}}} = U_{\mathrm{prep}} \ket{0}^{\otimes N+1}$, and the MPO representation of $S_2\!\left( \frac{\Delta t}{m} \right)^m$ with $m \gg 1$ is adopted as $e^{-iHt}$ in our case.
We define the approximate density matrix as
\begin{equation}
\begin{aligned}
    \mu(t) = \sum_i c_i(t)\, \rho_i(t),
    \label{eq: approximate density matrix}
\end{aligned}
\end{equation}
where  $\rho_i(t) = \ket{\psi_i(t)}\!\bra{\psi_i(t)}$, $\ket{\psi_i(t)} = \left(U_\mathrm{evol}^i\right)^{\frac{t}{\Delta t}} \ket{\psi_{\mathrm{prep}}}$, and $c_i(t)$ is a real coefficient for time $t$. The indexed operator $U_\mathrm{evol}^i$ can be prepared by changing the accuracy in the time evolution operator.

Here, we propose two procedures for preparing time evolution operators in $U_\mathrm{evol}^i$; the first one is changing the number of slices $m$ in the time evolution MPO in Eq.~\eqref{eq: reference time evolution} as
\begin{equation}
\begin{aligned}
    U_{\mathrm{ref}}^{i\dagger} &= S_2\!\left( \frac{-\Delta t}{m_i} \right)^{m_i}\\
    &=\left\{\prod_{\beta = 1}^{K} \exp (iC_{\beta} P_{\beta} \frac{\Delta t}{2{m_i}}) \prod_{\beta = K}^{1} \exp (iC_{\beta} P_{\beta}\frac{\Delta t}{2{m_i}})\right\}^{m_i}.
    \label{eq: approx time evol 1}
\end{aligned}    
\end{equation}
The second one is changing the number of sweeps in the optimization of $U_\mr{evol}$ in Fig.~\ref{fig: UnitaryCompression_1column.png}(c).

We minimize the squared Frobenius norm
\begin{equation}
\begin{aligned}
    &\norm{\rho_{\mathrm{p}}(t) - \mu(t)}_\mr{F}^2\\
    &= \Tr[\rho_{\mathrm{p}}(t)^{\dagger} \rho_{\mathrm{p}}(t)] + \Tr[\mu(t)^{\dagger} \mu(t)] - 2\Re\Tr[\rho_{\mathrm{p}}(t)^{\dagger}\mu(t)]\\
    &= 1 + \sum_{ij} M_{ij}(t)\, c_i(t)\, c_j(t)
      - 2 \sum_i L_i(t)\, c_i(t).
      \label{eq: aem norm}
\end{aligned}
\end{equation}
Here, the matrix elements are defined as
\begin{equation}
\begin{aligned}
    M_{ij}(t)
    &= \abs{
        \mel{\psi_{\mathrm{prep}}}{\left(U_\mathrm{evol}^{i\dagger}\right)^{\frac{t}{\Delta t}}
        \left(U_\mathrm{evol}^j\right)^{\frac{t}{\Delta t}}
        }{\psi_{\mathrm{prep}}}
      }^2 ,
\end{aligned}
\end{equation}
and
\begin{equation}
\begin{aligned}
    L_i(t)
    &= \abs{
        \mel{\psi_{\mathrm{prep}}}{
            e^{iHt} \left(U_\mathrm{evol}^i\right)^{\frac{t}{\Delta t}}
        }{\psi_{\mathrm{prep}}}
      }^2 .
\end{aligned}
\end{equation}

\begin{figure}[]
 \centering
 \includegraphics[width=1\columnwidth]{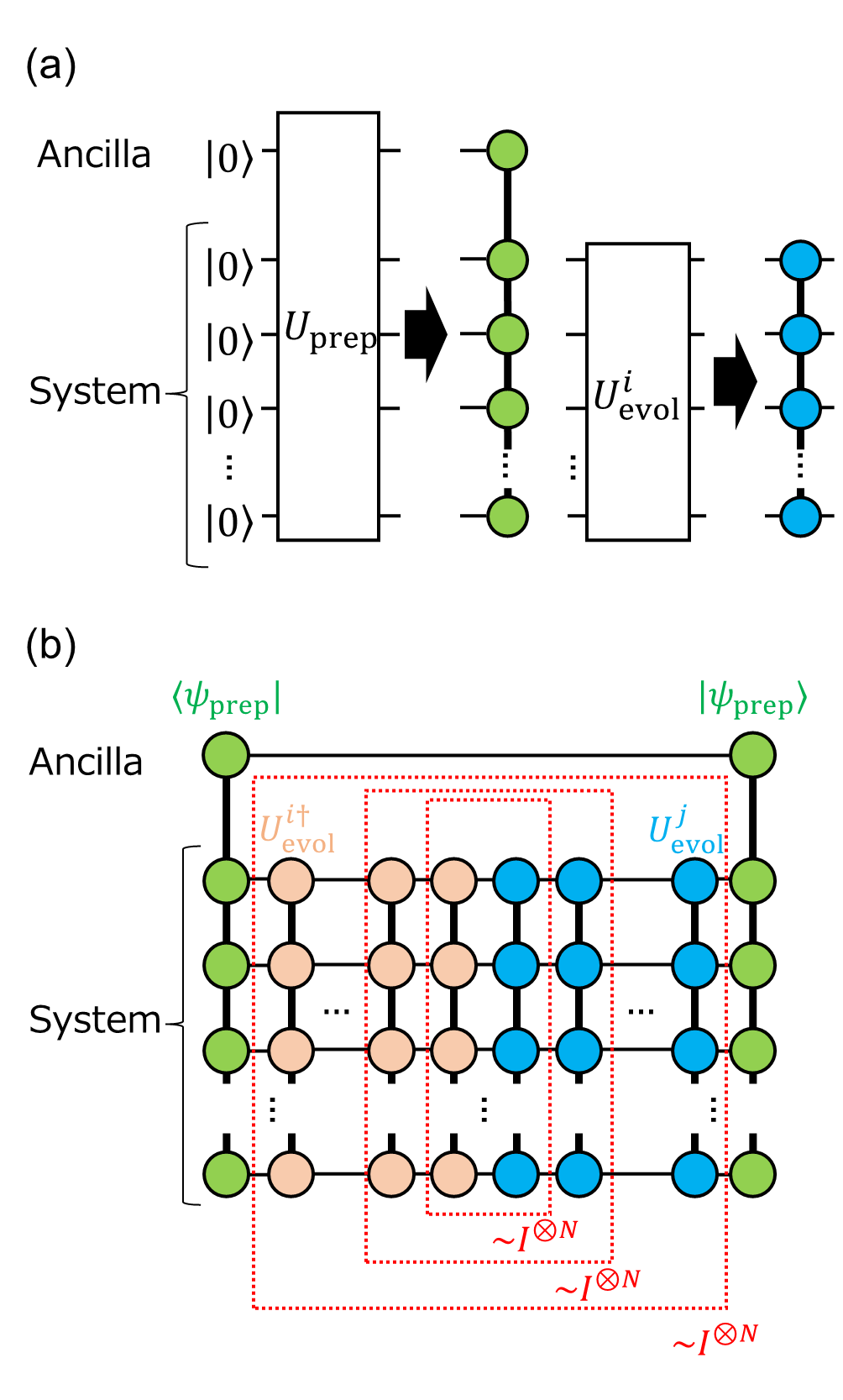}
\caption{Calculations for $M_{ij}(t)$ and $L_{i}(t)$. (a) Transformation from circuits of $\ket{\psi_\mr{prep}}$ and $U_\mr{evol}^i$ to the MPS and MPO, respectively. (b) Calculation procedure for $M_{ij}$. Note that $L_i$ can also be calculated in almost the same way.} 
 \label{fig: Mij_Li_1column.png}
\end{figure}
Since the objective function in Eq.~\eqref{eq: aem norm} is quadratic in \( c_i(t) \),
the optimal coefficients can be obtained if we calculate $M_{ij}(t)$ and $L_i (t)$. 
Note that we used constraints of $\sum_i |c_i(t)| \leq 3$ and $\sum_i c_i(t) = 1$ in the optimization of $c_i(t)$ on numerical demonstrations, where the latter corresponds a normalization of $\mu(t)$ since $\Tr[\mu(t)] = \sum_i c_i(t) \Tr[\rho_i(t)] = \sum_i c_i(t)$.

Figure~\ref{fig: Mij_Li_1column.png} shows our implementation for calculating $M_{ij}(t)$ and $L_i (t)$. We first transform $U_\mathrm{evol}^i$ and $U_\mathrm{prep}\ket{0}^{\otimes N+1}$ to the MPO and MPS, respectively, in Fig.~\ref{fig: Mij_Li_1column.png}(a). 
The MPS-MPO representation of $M_{ij}(t)$ is shown in Fig.~\ref{fig: Mij_Li_1column.png}(b).
Naively, the calculation of $(U_\mathrm{evol}^j)^{t/\Delta t}$ takes exponential classical cost since $(U_\mathrm{evol}^j)^{t/\Delta t} \approx e^{-iHt}$. 
To mitigate the exponential growth for $t$, we contract the network from the inside, as shown by the red dotted rectangles in Fig.~\ref{fig: Mij_Li_1column.png}(b), using the fact that $U_\mathrm{evol}^{i\dagger} U_\mathrm{evol}^j \approx I^{\otimes N}$~\cite{Robertson2024-jv}.
The computation of $L_i(t)$ can be carried out in the same manner by replacing $U_\mathrm{evol}^{i\dagger}$ with $e^{iH\Delta t}$.

The expectation value for $\mu(t)$ is represented using the optimized $c_i(t)$ as
\begin{equation}
    \begin{aligned}
    \Tr[O \mu(t)] &= \sum_i c_i (t) \Tr[O \rho_i (t)],
    \label{eq: refined expectation value}
    \end{aligned}
\end{equation}
where $O$ is an observable. So we can calculate $\Tr[O \mu(t)]$ by executing the circuits for $\Tr[O \rho_i (t)]$. 
We obtain the refined time series data $m(\theta)$ in Eq.~\eqref{eq: m(theta)} from Eq.~\eqref{eq: refined expectation value} by substituting $O = P(\theta)^\dagger U_{\mathrm{prep}} \ket{0}\bra{0}^{\otimes N+1}  U_{\mathrm{prep}}^\dagger P(\theta)$ and $\mu(t) = \rho(t)$.  
Note that in the substituted $O$ and $\mu(t)$, if $\rho_i (t)$ is affected by the deporizing noise independent on $i$, the value becomes $\Tr[O \mu'(t)] = (1 - p_{\mathrm{dep}}(t)) \sum_i c_i (t) \Tr[O \rho_i (t)] + \frac{p_{\mathrm{dep}}(t)}{2^{N+1}}$, almost the same expression as Eq.~\eqref{eq:m'(theta)}.

\begin{figure}[]
 \centering
 \includegraphics[width=1\columnwidth]{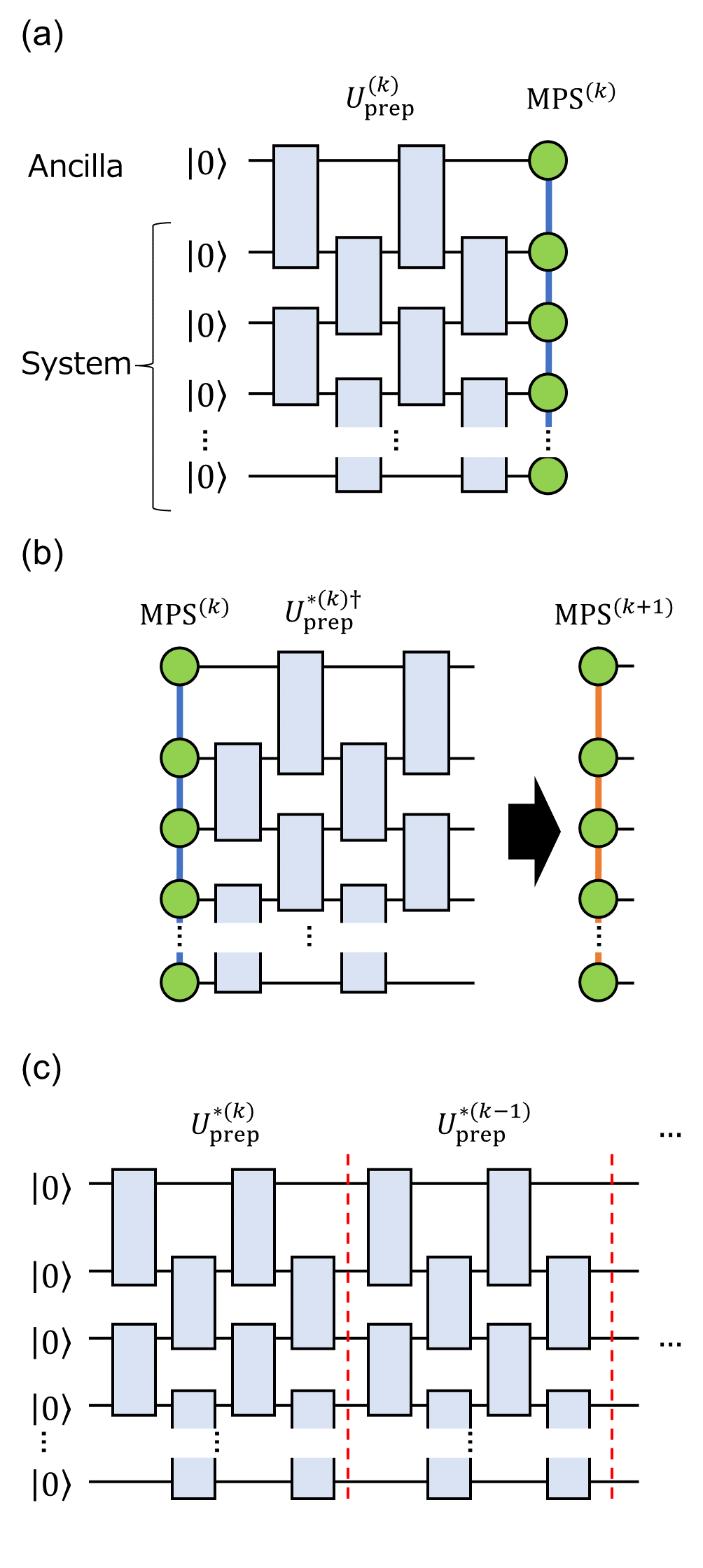}
\caption{Procedure of the overlap enhancement. (a) Circuit optimization. (b) Circuit merging. (c) Deep circuit implementation.} 
 \label{fig: overlap_enhancement_explain_1column.png}
\end{figure}

\subsection{Overlap enhancement for a compressed state preparation circuit}
In the circuit optimization described in Sec.~\ref{sec: summary of classical unitary compression using tensor networks}, increasing the depth $d_{\mathrm{prep}}$ allows $U_{\mathrm{prep}}$ to prepare a state closer to  $\ket{\mathrm{MPS}}$ in Eq.~\eqref{eq: prep}. 
However, the optimization time also grows exponentially with respect to $d_{\mathrm{prep}}$ due to the increase of the bond dimension in the MPS. 
The cost of contracting the tensor network is $\order{\chi d 2^{4+2\ceil{d/2}}+\chi^2 2^{2+2\ceil{d/2}}}$, where $d \in \{d_\mr{prep}, d_\mr{evol}\}$, and $\chi$ is the maximum bond dimension in the reference MPOs~\cite{Causer2024-wd}.

In reality, in the 37-qubit circuit for the one-dimensional Hubbard model presented later, performing 1000 sweeps requires only several tens of minutes on a consumer model laptop computer when $d_{\mathrm{prep}} = 5$, whereas it takes more than a week on our high-performance computing system when $d_{\mathrm{prep}} = 12$. 
The overlap was 0.97 for the 9-qubit circuit at $d_{\mathrm{prep}} = 5$, whereas for the 37-qubit circuit it was only 0.76 even at $d_{\mathrm{prep}} = 12$. Such degradation of the overlap generally leads to an increase in computational cost. For example, the cost of ground-state estimation via time-series QPE scales as $\eta^{-4}$ with respect to the absolute value of overlap $\eta$ between the initial state and the exact ground state~\cite{Lin2022-rn}. Therefore, it is necessary to prepare a more accurate wavefunction using deeper circuits while avoiding exponential growth in the computational cost.

We propose a technique to alleviate this problem via iterative optimization. The procedure is as follows and shown in Fig.~\ref{fig: overlap_enhancement_explain_1column.png}.
\begin{enumerate}
\item We set the index $k=1$.
\item As in Fig.~\ref{fig: overlap_enhancement_explain_1column.png}(a), $U_\mr{prep}^{(k)}$ is optimized to maximize the real part of overlap $\Re\left[\mel{\mr{MPS}^{(k)}}{U_\mr{prep}^{(k)}}{0}^{\otimes N+1}\right]$, where we added the index $k$ for $\ket{\mr{MPS}}$ and $U_\mr{prep}$ to identify the iteration count. The state $\ket{\mr{MPS}^{(1)}}$ is the target MPS that we want to approximate.
\item As in Fig.~\ref{fig: overlap_enhancement_explain_1column.png}(b), using the optimized circuit $U_\mr{prep}^{*(k)}$, we create $\ket{\mr{MPS}^{(k+1)}} = U_\mr{prep}^{*(k)\dagger} \ket{\mr{MPS}^{(k)}}$ which is closer to $\ket{0}^{\otimes N+1}$ than $\ket{\mr{MPS}^{(k)}}$.
\item We increase $k \rightarrow k+1$ for $U_\mr{prep}^{(k)}$. The procedures 2-4 are repeated until a termination condition is satisfied.
\item As in Fig.~\ref{fig: overlap_enhancement_explain_1column.png}(c), we obtain a deep circuit $U_{\mr{prep}} = \prod_{k=k_\mr{max}}^1 U_\mr{prep}^{*(k)}$.
\end{enumerate}
$U_\mr{prep}$ give a higher overlap than single $U_{\mr{prep}}^{(1)}$ since 
\begin{equation}
    \begin{aligned}
    \mel{\mr{MPS}^{(1)}}{U_\mr{prep}}{0}^{\otimes N+1} &=  \mel{\mr{MPS}^{(1)}}{\prod_{k=k_\mr{max}}^1 U_\mr{prep}^{*(k)}}{0}^{\otimes N+1}\\
    &= \mel{\mr{MPS}^{(2)}}{\prod_{k=k_\mr{max}}^2 U_\mr{prep}^{*(k)}}{0}^{\otimes N+1}\\
    &= \dots\\
    &= \braket{\mr{MPS}^{(k_\mr{max})}}{0}^{\otimes N+1}.
    \end{aligned}
\end{equation}
Note that the bond dimension of $\ket{\mr{MPS}^{(k)}}$ would increase for $k$ in general, but it may decrease in a high overlap regime since
when $\Re\left[\mel{\mr{MPS}^{(k)}}{U_\mr{prep}^{*(k)}}{0}^{\otimes N+1}\right] \approx 1$, $\ket{\mr{MPS}^{(k+1)}} = U_\mr{prep}^{*(k)\dagger} \ket{\mr{MPS}^{(k)}} \approx \ket{0}^{\otimes N+1}$, i.e., a separable state.
Also, a related approach has been proposed, in which, instead of maximizing the overlap with the MPS, each gate is optimized and merged into the MPS to reduce the entanglement entropy of the MPS~\cite{Mansuroglu2026-ay}. The gate optimization is executed only once for each gate in the approach, so it can be useful when prioritizing the improvement of the overlap over circuit-depth efficiency.

\section{Numerical demonstration}
\label{sec: numerical demonstration}

\subsection{Models and calculation conditions}
\label{sec: models}
We execute numerical experiments using a noiseless simulator and real devices. We estimate the energy gap between the ground state and the first excited state energies, which is represented as $\Delta_{(0,1)}^\mr{est}$. The target system is the one-dimensional Hubbard model:
\begin{equation}
\begin{aligned}
H &= -T \sum_{q=1}^{n_{\mathrm{s}}-1} \sum_{\sigma \in \{\uparrow, \downarrow \}}(a_{q+1\sigma}^{\dagger} a_{q\sigma}+a_{q\sigma}^{\dagger} a_{q+1\sigma})\\
&+U \sum_{q=1}^{n_{\mathrm{s}}} n_{q\uparrow} n_{q\downarrow}
-\frac{U}{2} \sum_{q=1}^{n_{\mathrm{s}}} (n_{q\uparrow}+n_{q\downarrow}),
\label{Eq: Hubbard model}
\end{aligned}
\end{equation}
where $q$ ($\sigma$) is the orbital (spin) index, $T=1$ is the unitless hopping energy, $U$ is the on-site Coulomb energy, $a_{q\sigma}^{\dagger}$ ($a_{q\sigma}$) is the creation (annihilation) operator, and $n_{q\sigma}$ is the number operator $n_{q\sigma} = a_{q\sigma}^{\dag} a_{q\sigma}$. 
We choose $U=10$ corresponding to the strong correlation regime, the number of system qubits $N$ is $2 n_{\mr{s}}$, where $n_{\mr{s}}$ is the number of physical sites, and the fermion-qubit mapping is a Jordan-Wigner transformation~\cite{Jordan1928-hf} with an up-down qubit sequence $1\uparrow, 1\downarrow, 2\uparrow,\dots,n_{\mathrm{s}}\downarrow$.

Our default conditions are as follows. 
We calculate the circuits of the $N=8,36,52$ system qubits with one ancilla qubit.
The value of $d_\mr{prep} = 5$ for the 9-qubit circuit and $12$ for the 37- and 53-qubit ones, and $d_\mr{evol} = 5$. The number of sweeps is $10^3$ for $U_\mr{prep}$. For $U_\mr{evol}$, the number is $10^4$. The cutoff value of the SVD in the optimization of $U_\mr{prep}$ and $U_\mr{evol}$ is $10^{-12}$. The number of slices $m$ in Eq.~\eqref{eq: reference time evolution} is $\Delta t / 0.001$. The number of shots in a circuit execution is $10^5$. The maximum number of time steps is 100 in the next section and 50 the others, but the execution is stopped when no signal is found for a time step. 

We calculate reference energies for the gap $\Delta_{(0,1)}^{\mr{ref}}$ using the exact diagonalization for the 8-qubit model and DMRG with maximum bond dimension 1,000 for the other large models.
The tensor network compression was executed by ITensor~\cite{Fishman2020-lw}, and the circuits were executed by Qiskit~\cite{Javadi-Abhari2024-qm}.
We use the Q-CTRL error suppression module, Fire Opal~\cite{Mundada2023-rl, Baum2021-sd}, for real device demonstrations.
We used the Aer simulator for the noiseless simulations and IBM Heron devices for real device execution.

The other conditions are the same as the previous work~\cite{Kanno2025-tt}.

\begin{figure}[]
 \centering
 \includegraphics[width=1\columnwidth]{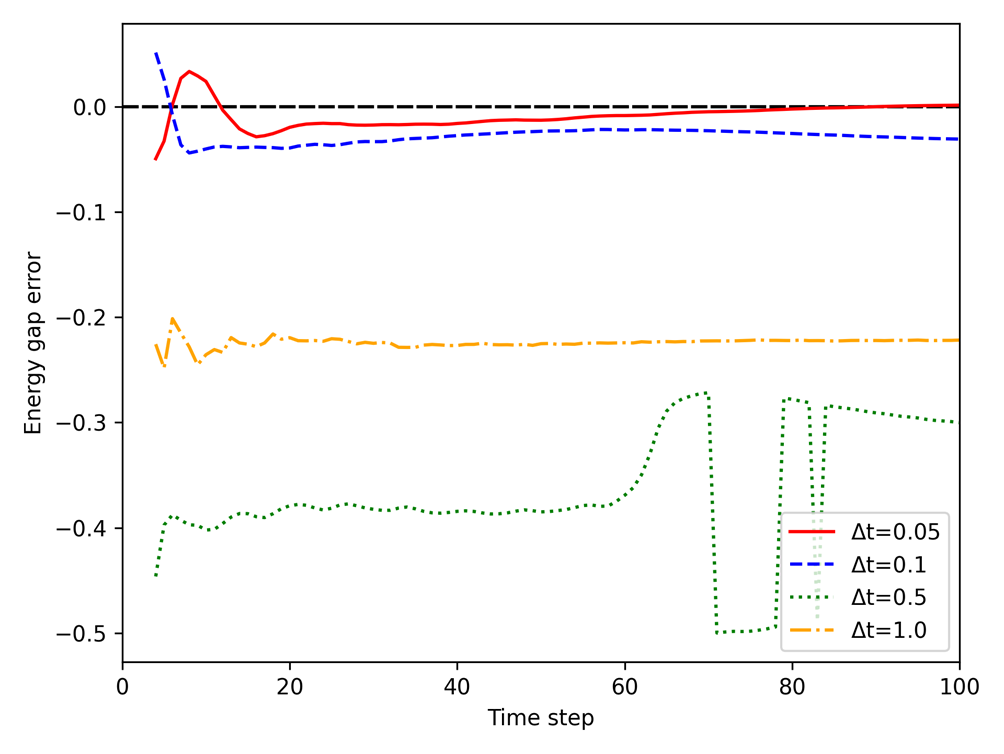}
\caption{Energy gap errors, $\Delta_{(0,1)}^\mr{est} - \Delta_{(0,1)}^\mr{ref}$, for time step with several step sizes $\Delta t$.} 
 \label{fig: dt_dependency_1column.png}
\end{figure}

\subsection{Time step size dependencies}
\label{sec: time step depend}
We confirmed the accuracy of the time step size $\Delta t$ in the nine-qubit circuit. 
Figure~\ref{fig: dt_dependency_1column.png} shows the error of the estimated gap from the reference gap for each time step, $\Delta_{(0,1)}^\mr{est} - \Delta_{(0,1)}^\mr{ref}$, where the exact gap $\Delta_{(0,1)}^\mr{ref}$ is 0.254, and the black dashed horizontal line represents the zero gap error. The gap at a given time step was computed using the time-series data up to that point. Note that because the step size differs between plots, the total time $t$ corresponding to the same step number is not identical.

For $\Delta t = 0.1$ and $0.05$, the final absolute error was below $0.1$.
Excluding $\Delta t = 0.5$, the fluctuations stabilize around step 50 regardless of $\Delta t$.
For $\Delta t = 0.5$, the results became unstable after 60 steps, likely due to the accumulation of unitary-approximation errors.
Further considering that large-scale hardware demonstrations may not reliably reach 100 steps, we use 50 steps in the analysis from the next section.

Regarding the time-step size, noiseless simulations are performed with $\Delta t = 0.05$ in the following section, where the absolute relative errors at 50 and 100 steps were 0.012 and 0.001, corresponding to the error ratios of 4.7\% and 0.4\%, respectively.
Here, we aim for target accuracy as $T/100=0.01$~\cite{Kivlichan2020-sp}.
On the other hand, we set $\Delta t = 0.1$ on real device demonstrations since Fire Opal provided the best balance between algorithmic precision and hardware noise in this value.

\begin{figure}[]
 \centering
 \includegraphics[width=1\columnwidth]{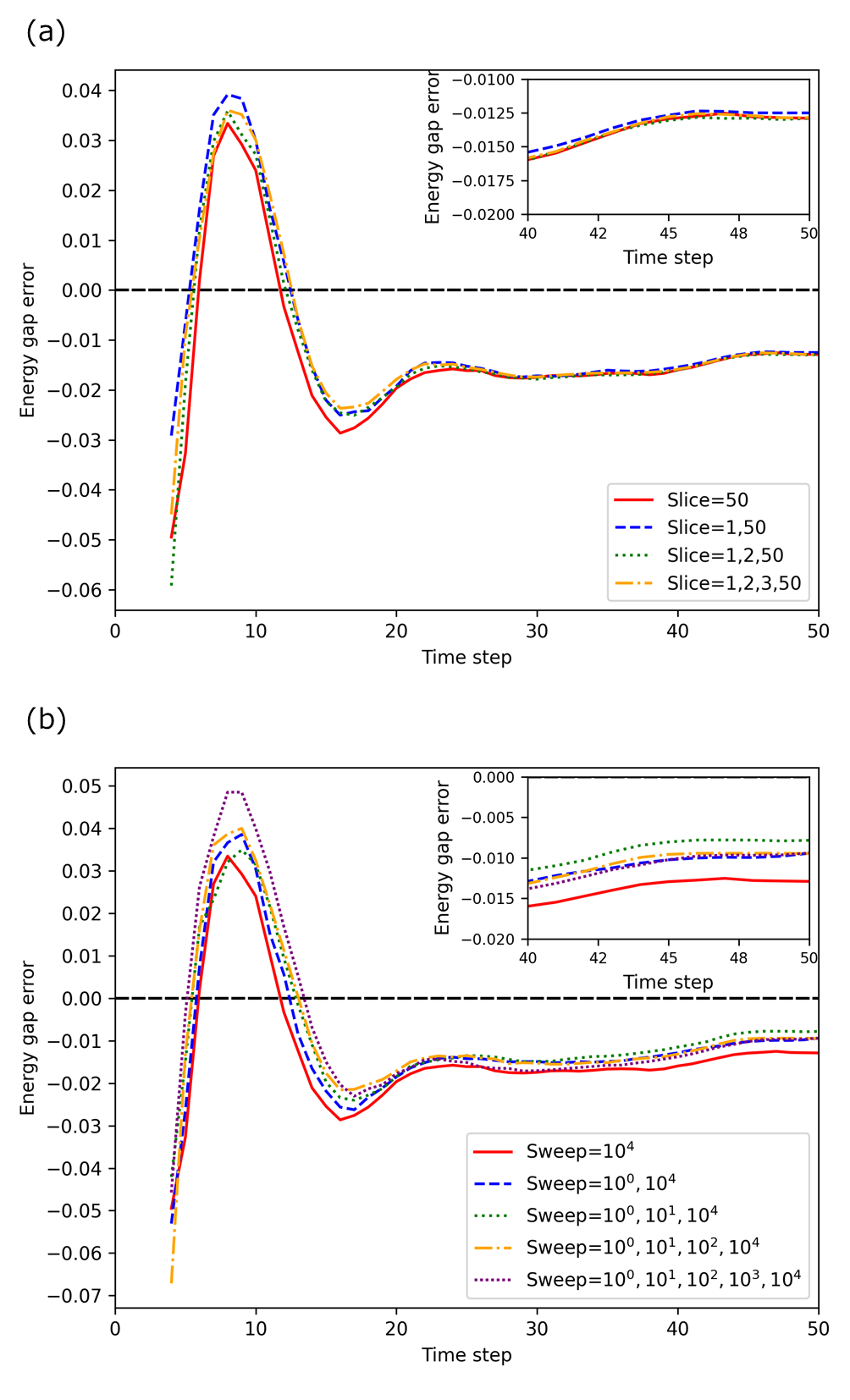}
\caption{Energy gap errors, $\Delta_{(0,1)}^\mr{est} - \Delta_{(0,1)}^\mr{ref}$, with and without the AEM. The inset shows an enlarged view. (a) Dependency on slice sets. (b) Dependency on sweep sets.} 
 \label{fig: step_dependency_1column.png}
\end{figure}

\begin{figure*}[]
 \centering
 \includegraphics[width=1\textwidth]{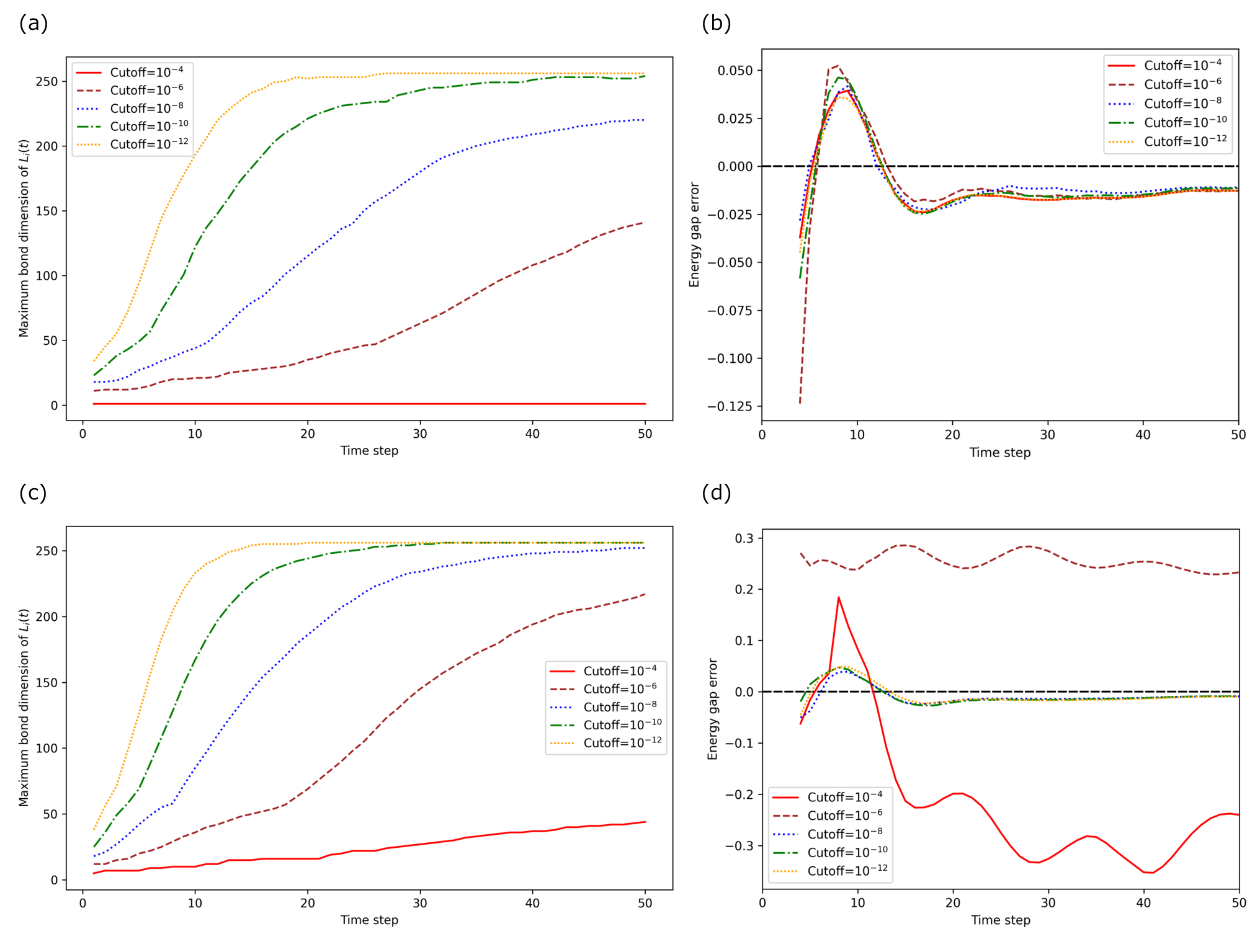}
\caption{SVD dependencies for the bond dimension and energy gap errors, $\Delta_{(0,1)}^\mr{est} - \Delta_{(0,1)}^\mr{ref}$. (a) and (b) [(c) and (d)] are the plots for the slice [sweep] case. (a) and (c) represent the maximum bond dimensions for $L_{i}(t)$, and (b) and (d) represent the gap for the time step.} 
 \label{fig: svd_dependency_2column.png}
\end{figure*}

\subsection{Verification of algorithmic error mitigation}
\label{sec: verify AEM}
We verify the AEM for the two cases of preparation: changing the number of slices/sweeps.
Figure~\ref{fig: step_dependency_1column.png} (a) and (b) show the slice and the sweep dependencies for the gap in the $N+1=9$ qubit circuit. We do not find a significant improvement for the slice case. On the other hand, we do find the improvement in the sweep case. The degree of improvement depends on the included sweep sets, but the absolute error becomes less than 0.01 for all of the sweep lists.

We can reduce the calculation cost by choosing rough cutoffs for SVD in the calculation procedure of $M_{ij}(t)$ and $L_i(t)$.
Figure~\ref{fig: svd_dependency_2column.png} (a) and (c) show the maximum bond dimension of $L_i(t)$, precisely that of $e^{iHt} (U_\mathrm{evol}^i)^{t/\Delta t}$, for the time step in the slice and sweep cases, respectively. The values rapidly increase for the step and saturate the maximum value 256, but the growth is suppressed by choosing the rough cutoffs, where the maximum bond dimension for the $N+1$ qubit MPO is $4^{\lfloor \frac{N+1}{2} \rfloor}$~\cite{Schollwock2011-im}.
Figure~\ref{fig: svd_dependency_2column.png} (b) and (d) show the cutoff dependencies for the gap for the time step in the slice and sweep cases, respectively. 
From the sweep case, the AEM would correctly work up to the cutoff value of $10^{-8}$.

\begin{figure}[]
 \centering
 \includegraphics[width=1\columnwidth]{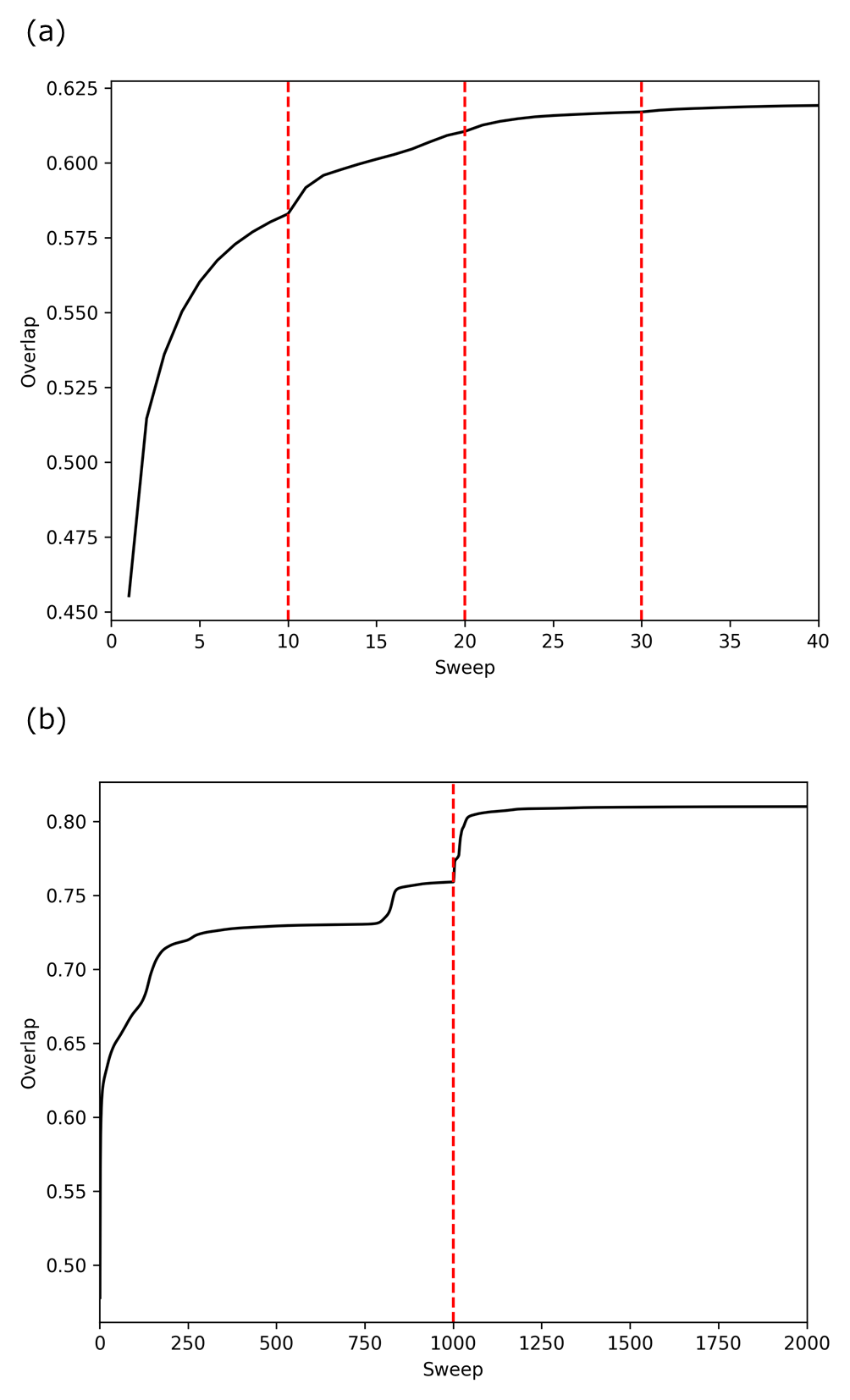}
\caption{Overlap enhancement. The vertical break line in red represents the increment of $k$. (a) Verification for $d_\mr{prep}/iter = 6$ with 10 sweeps for each $k$ (b) Verification for $d_\mr{prep}/iter = 12$ with 1000 sweeps for each $k$.} 
 \label{fig: overlap_enhancement_1column.png}
\end{figure}

\subsection{Verification of overlap enhancement}
\label{sec: verification of overlap enhancement}
We evaluated the overlap enhancement for a circuit with $N+1 = 37$ qubits, where the 9-qubit circuit was not tested since its accuracy was already sufficient at $d_{\mathrm{prep}} = 5$. We first performed test calculations with the depth increment per iteration set to $d_{\mathrm{prep}}/iter = 6$, the number of sweeps for each optimization of $U_{\mathrm{prep}}^{(k)}$ set to 10, and the SVD cutoff during optimization set to $10^{-8}$. Figure~\ref{fig: overlap_enhancement_1column.png}(a) shows the evolution of the overlap as a function of the sweep count; the dashed vertical lines indicate where the iteration index is incremented. The overlap increases as the number of sweeps grows, and the procedure successfully ran up to $d_{\mathrm{prep}} = 24$ without running out of memory, which is not achievable with the previous method. However, the final overlap 0.62 was much smaller than the value obtained at $d_{\mathrm{prep}} = 12$ in the previous approach, 0.76. This indicates that unless $d_{\mathrm{prep}}/iter$ and the sweep count are set sufficiently large, increasing the circuit depth does not yield a significant overlap improvement.

The results for the case with $d_{\mathrm{prep}} / \mathit{iter} = 12$ and 1000 sweeps for each $U_{\mathrm{prep}}^{(k)}$ are shown in Fig.~\ref{fig: overlap_enhancement_1column.png}(b). The calculation ran out of memory at $\mathit{k}=2$. The resulting overlap was 0.81, which is larger than 0.76. These results demonstrate that, with appropriate choices of circuit depth per iteration and sweep conditions, it is possible to prepare states with larger overlap than in the previous approach.

\begin{figure*}[]
 \centering
 \includegraphics[width=1\textwidth]{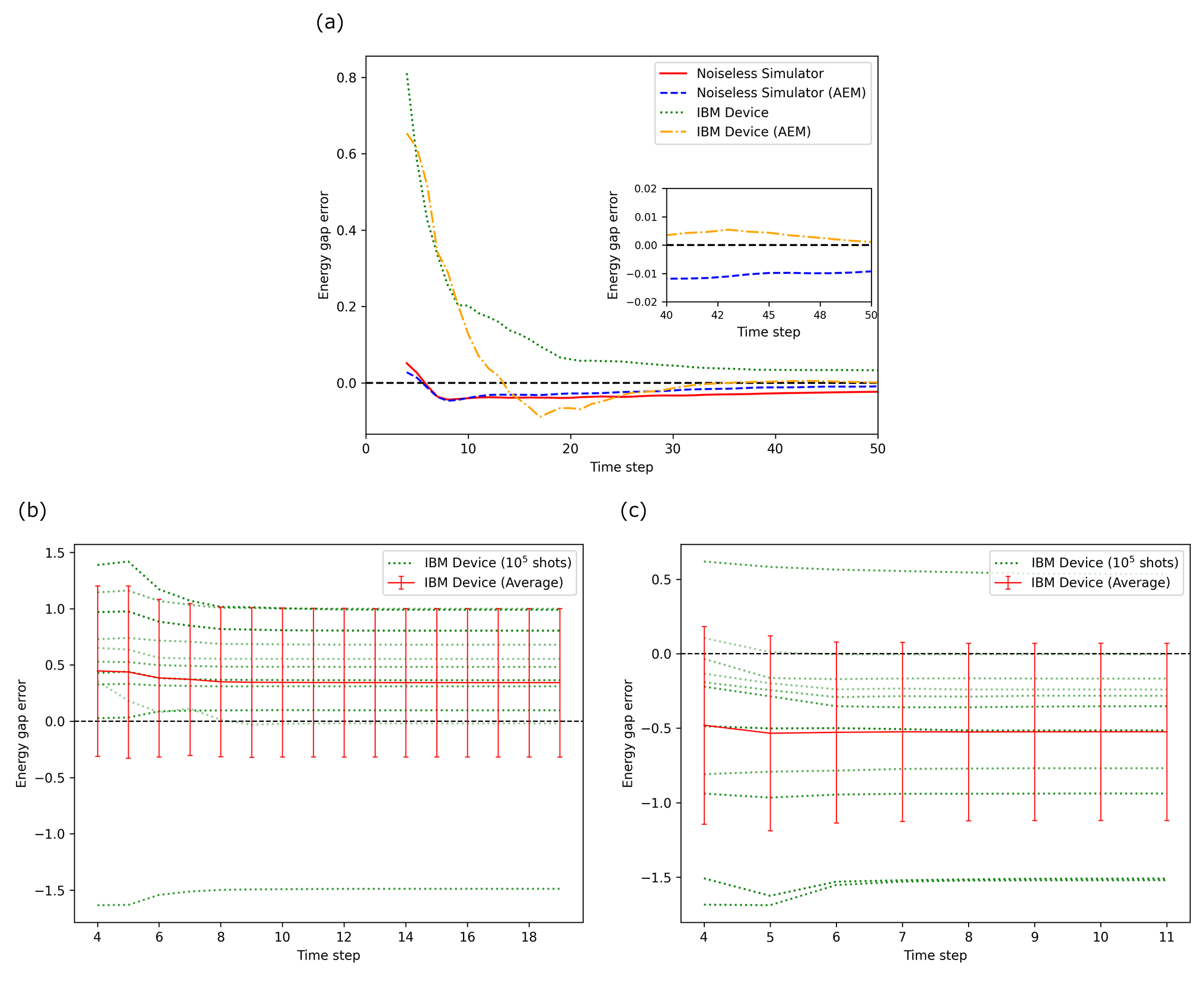}
\caption{Real device demonstrations for energy gap errors, $\Delta_{(0,1)}^\mr{est} - \Delta_{(0,1)}^\mr{ref}$ using \textit{ibm\_kawasaki}. The inset in (a) shows an enlarged view. The error bars in (b) and (c) represent standard deviations, and 11 trials are represented by gradated green dotted lines.  (a) 9-qubit circuit results, including noiseless simulator results) (b) 37-qubit circuit results (c) 53-qubit circuit results.} 
 \label{fig: real_device_2column.png}
\end{figure*}

\subsection{Real device demonstration}
\label{sec: real device demonstration}
Figure~\ref{fig: real_device_2column.png}(a) shows the results of gap estimation for the 9-qubit circuit with $\Delta t = 0.1$. From the noiseless simulator results, we find that even at $\Delta t = 0.1$, higher accuracy is achieved with the AEM (blue) than without the AEM (red). Here, the AEM uses circuits with sweep counts of $10^0$, $10^1$, and $10^4$. The obtained values are $-0.023$ and $-0.009$, respectively, indicating that the use of the AEM yields an accuracy on the order of $10^{-3}$. Similar behavior is observed on real device demonstration, where the result with the AEM (yellow) also achieves a high accuracy of 0.001, which is better than without the AEM of 0.033 (green).
The numbers of 2-qubit (CZ and Rzz) gates in the final steps with and without the AEM are 1699 and 1698, respectively.

Figures~\ref{fig: real_device_2column.png}(b) and (c) show the results for the 37 and 53 qubit circuits, respectively. 
We executed 11 trials since the large variances were expected due to rapid signal decay: only up to 19 and 11 steps, respectively, were executed before the signal was lost.  
The errors were in the order of 0.1 for averages (red lines), while below the order of 0.01 or less in the best trials, where the reference gap values $\Delta_{(0,1)}^\mr{ref}$ are 0.076 and 0.055, mean gap errors are 0.343 and -0.524, and the numbers of 2-qubit gates were 4146 and 4091 for 37 and 53 qubit models, respectively.
The zero errors lie within the error bars, where the standard deviations are 0.660 and 0.595 for 37 and 53-qubit models, respectively. 

We mention that the classical computation in the AEM, specifically, the evaluation of $L_i(t)$, was halted at 11 steps for those systems, where the suppression of the cost increase 
by the implementation shown in Fig.~\ref{fig: Mij_Li_1column.png}(b) is discussed in Appendix~\ref{sec: bond dimension for real device executions}.
Although an efficient implementation of AEM may be needed, from the viewpoint of scaling, it may be possible to achieve a similar scaling for the accuracy in time evolution even without using AEM. See Appendix~\ref{sec: depth scaling} for the analysis.
We also mention that hardware execution was attempted for the 37-qubit model with $d_{\mathrm{prep}}/\mathit{iter} = 2$ and $d_{\mathrm{prep}} = 24$, but the run terminated after four steps, where the gate count was 2684.

From these results, our algorithm demonstrates not only high performance in large-scale QPE-type algorithms but also highlights challenges that remain for the future: the proposed method has been shown to be an efficient approach well-suited for current devices. With future improvements in gate performance, it is expected that the large-scale demonstration for QPE-type algorithms can be executed with higher precision.

\section{conclusion}
\label{sec: conclusion}
In this study, we proposed a phase-estimation-type algorithm that extracts time-series data related to the energy gap from quantum circuits compressed using tensor networks. The circuits considered here use only a single ancilla qubit and system qubits and are composed of only neighboring gates. By combining measurement results from four types of circuits, the time-series data represented as complex numbers can be obtained. Numerical validation shows that, with an appropriate choice of the time-step size, the results converge within 50 steps regardless of the time steps. In addition, we proposed techniques to improve the accuracy of both time evolution and state preparation in our algorithm. 
The former is based on the AEM, while the latter relies on iterations combining circuit optimization with the MPS merging. We executed the time series algorithm on noiseless devices and IBM quantum hardware using Q-CTRL's error-suppression module. For one-dimensional Hubbard models with 8, 36, and 52 qubits, accuracies on the order $10^{-3}$ or larger were observed depending on conditions. 

Future work may include achieving higher accuracy and larger-scale implementations by developing more efficient algorithms on both the quantum and classical sides. For example, tensor networks other than an MPS could be employed~\cite{D-Anna2025-wk, Sugawara2025-xu, Gibbs2025-hd, Karacan2026-zh}. 
In addition, the reduction of classical computational overhead is important since the AEM and the quantum computational cost associated with improving state-preparation accuracy are both large.
Also, there is a possibility that the sampling cost can be reduced by modifying the protocol. 
In this work, from the standpoint that it is difficult to estimate the maximum executable time $t_\mr{max}$ and the decay of the signal $s_t$ in advance, we obtained $s_t$ for sequential values from $t=1$ to $t_\mr{max}$. 
If the decay can be estimated easily, or if there is no decay due to error correction, it may be possible to use a more efficient protocol that achieves the Heisenberg limit~\cite{Lin2022-rn, Ding2023-iu} by adaptive strategies.
Finally, since this algorithm is crucial not only for the current devices but also for FTQC, verifications for QPE focusing on the FTQC era are also promising.

\section{acknowledgment}
A part of this work was performed for Council for Science, Technology and Innovation (CSTI), Cross-ministerial Strategic Innovation Promotion Program (SIP), “Promoting the application of advanced quantum technology platforms to social issues”(Funding agency: QST). 
Also, it was supported by MEXT Quantum Leap Flagship Program Grants No. JPMXS0118067285 and No. JPMXS0120319794. 
The part of the calculations was performed on the Mitsubishi Chemical Corporation (MCC) high-performance computer (HPC) system “NAYUTA”, where “NAYUTA” is a nickname for MCC HPC and is not a product or service name of MCC.
We acknowledge the use of IBM Quantum services for experiments in this paper. The views expressed are those of the authors and do not reflect the official policy or position of IBM or the IBM Quantum team.
Thank you for the support of the error suppression module by Q-CTRL.
S.K. thanks Keithley Kimberlee for careful reading of the manuscript and for helpful suggestions on English usage. K.S. acknowledges partial support from Center of Innovations for Sustainable Quantum AI (JPMJPF2221) from Japan Science and Technology Agency (JST), Japan and Grants-in-Aid for Scientific Research C (21K03407) and for Transformative Research Area B (23H03819) from Japan Society for the Promotion of Sciences (JSPS), Japan.

\section{Code and data availability}
The codes and datasets in our study can be available at Ref.~\cite{UnknownUnknown-mg}.  

\appendix
\section{Eigenvalue extraction from time series data}
\label{sec: eigenvalue extraction from time series data}
Assuming that the signal \( s_t \) is obtained from the quantum circuits as
\begin{equation}
\begin{aligned}
    s_t = e(t)\sum_J P_J e^{-i \Delta_J t} ,
\end{aligned}
\end{equation}
where \(e(t)\) is a signal attenuation, and the sampling times are \( t \in \{ t_r\} (t_r = t_1, t_2,\dots, t_\mr{max}) \) with fixed intervals $\Delta t = t_{i+1} - t_i$. The signal attenuation can be taken as $e(t) \approx e^{-\alpha t}$ under depolarizing noise, although it depends on circuit compilation details.

We adopt the matrix pencil method as an initial guess.
We construct the shifted matrices
\begin{equation}
\begin{aligned}
    A_0 &=
    \begin{pmatrix}
        s_1 & s_2 & \cdots & s_l \\
        s_2 & s_3 & \cdots & s_{l+1} \\
        \vdots & \vdots & \ddots & \vdots \\
        s_{m-1} & s_m & \cdots & s_{m+l-2}
    \end{pmatrix},
\end{aligned}
\end{equation}
and
\begin{equation}
\begin{aligned}
    A_1 =
    \begin{pmatrix}
        s_2 & s_3 & \cdots & s_{l+1} \\
        s_3 & s_4 & \cdots & s_{l+2} \\
        \vdots & \vdots & \ddots & \vdots \\
        s_m & s_{m+1} & \cdots & s_{m+l-1}
    \end{pmatrix},
\end{aligned}
\end{equation}
where we choose $l=t_\mr{max}/2$ and $m=t_\mr{max}/2 +1$.
The eigenvalues are obtained by solving the generalized eigenvalue problem
\begin{equation}
\begin{aligned}
    A_1 \, x = \lambda \, A_0 \, x .
\end{aligned}
\end{equation}
The eigenvalue corresponding to the dominant spectral component is expressed as
\begin{equation}
\begin{aligned}
    \lambda \approx g e^{-(i \Delta_J + \alpha)\Delta t},
\end{aligned}
\end{equation}
and $\Delta_J$ corresponds to the desired eigenvalue and $g$ is a constant. The amplitude $P_{J}$ is determined by fitting $g$ as 
\begin{equation}
    \begin{aligned}
        \min_{P_J} \sum_t | s_t - P_{J} e^{-(i \Delta_J + \alpha)\Delta t}|^2,
    \end{aligned}
\end{equation}
using the least squares method.

Then we refine the eigenvalue by non-linear simultaneous optimization of $P_J$, $\Delta_J$, and $\alpha$ (i.e., data fitting) using the initial guess. 
\begin{equation}
    \begin{aligned}
        \min_{P_J, \Delta_J, \alpha} \sum_t | s_t - P_{J} e^{-(i \Delta_J + \alpha)\Delta t}|^2,
    \end{aligned}
\end{equation}
where we used the least squares method for the optimization.

\section{Bond dimension for real device executions \label{sec: bond dimension for real device executions}}
\begin{figure}[]
 \centering
 \includegraphics[width=1\columnwidth]{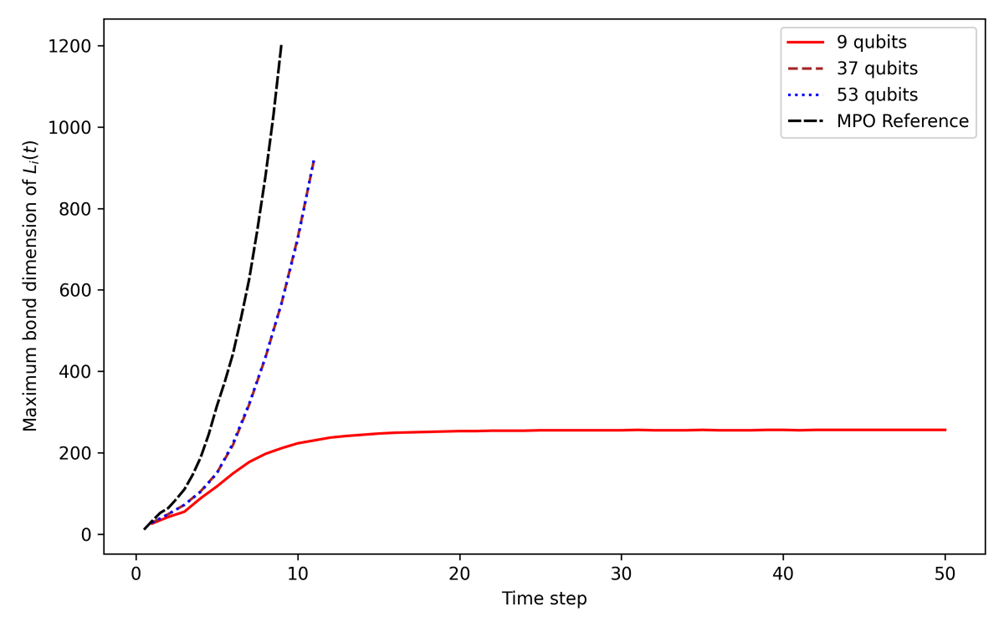}
\caption{SVD dependencies for the bond dimension in $\Delta t = 0.1$. The calculations for $M_{ij}$ and $L_i$ are executed using the sweep of $10^0$, $10^1$, and $10^4$ counts. The reference is the value for MPO of $(U_{\mathrm{ref}}^\dagger)^{\frac{t}{2\Delta t}}$ in the 52 qubit system.} 
 \label{fig: SVD_rd_AEM.png}
\end{figure}

Figure ~\ref{fig: SVD_rd_AEM.png} shows the SVD dependencies for the bond dimension in Fig.~\ref{fig: real_device_2column.png}. 
The value in the 9-qubit circuit (red) was saturated at 256. The values in the larger models (brown for 37- and blue for 53-qubit circuits) were increased. The values are almost the same for the 37- and 53-qubit circuits, indicating that the increase is independent of size for large systems. 
We also show the values for the MPO of $(U_{\mathrm{ref}}^\dagger)^{\frac{t}{2\Delta t}}$ in 52 qubits for reference, where $\frac{t}{2\Delta t}$ (half time step value) comes from the fact that two MPOs are merged in the calculation for $M_{ij}(t)$ and $L_i(t)$ in Fig.~\ref{fig: Mij_Li_1column.png}(b), and the SVD cutoff is $10^{-8}$.
We found that the increase in calculation cost is suppressed in the AEM due to the merge.
Considering that the bond dimension more than $10^{4}$ is tractable by a high-end HPC system~\cite{Ganahl2023-nh}, it is expected that the use of more powerful classical devices or efficient contraction schemes may enable the AEM to be executed over longer durations.

\section{Depth scaling for compressed time evolution circuit \label{sec: depth scaling}}
\begin{figure}[]
 \centering
 \includegraphics[width=1\columnwidth]{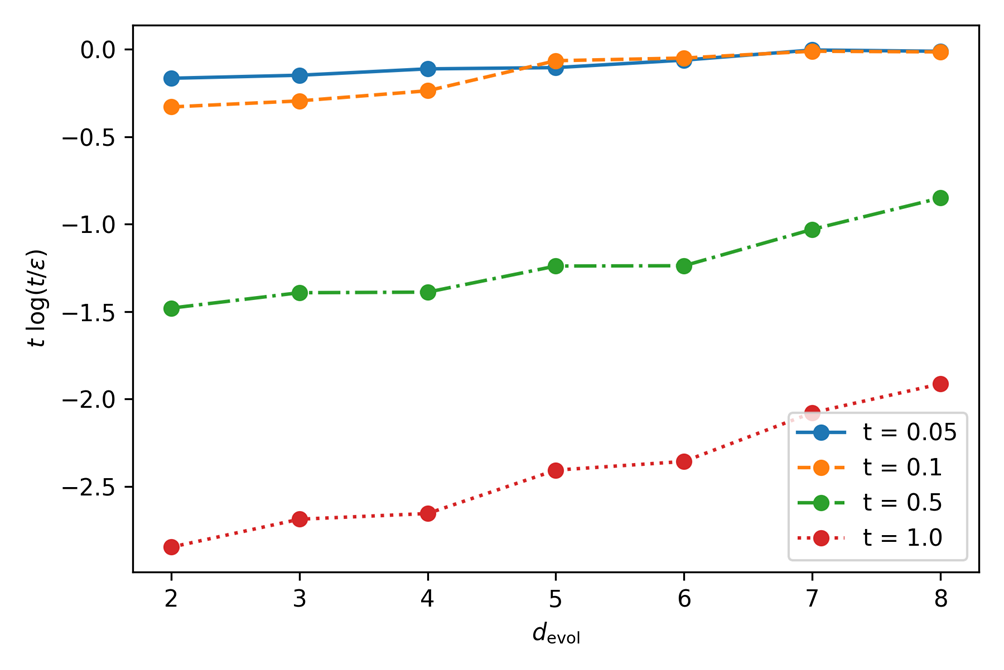}
\caption{Scaling for the depth $d_\mr{evol}$ in the Hubbard model. We adopt $\Delta t$ and $\norm{U_{\mathrm{ref}}^\dagger - U_{\mathrm{evol}}}_\mr{F}$ as $t$ and $\varepsilon$ in the compressed circuits, respectively.} 
 \label{fig: scaling_1column.png}
\end{figure}
We analyze the depth scaling of $d_{\mathrm{evol}}$ in the one-dimensional Hubbard model.
AEM was introduced to mitigate Trotter errors~\cite{Endo2019-oh}, reducing the order of the error in $p$-th order Trotter approximation from $\order{t^{1+\frac{1}{p}} /\varepsilon^{\frac{1}{p}}}$ to the near-optimal scaling for $\varepsilon$, $\order{t~\mr{polylog}(t/\varepsilon)}$ without relying on the coherent techniques~\cite{Watson2024-bn, Mizuta2026-yl}.
On the other hand, besides AEM, there exists a procedure to decompose the time evolution operator of a short-range Hamiltonian into gates with circuit depth $\order{t~\mr{polylog}(t/\varepsilon)}$ using the Lieb–Robinson bound~\cite{Haah2021-wm}. In the related work~\cite{Karacan2026-zh}, it has been numerically suggested that a tensor-compressed time evolution circuit can also be implemented with this scaling, where a one-dimensional Heisenberg model was adopted.

We therefore examined whether this scaling holds for the one-dimensional Hubbard model as well, based on a procedure in Ref.~\cite{Karacan2026-zh}. For several values of $\Delta t$, we computed $U_{\mathrm{evol}}$ while varying $d_{\mathrm{evol}}$. To confirm the dependence of $d_{\mr{evol}} \sim t \log (t/\varepsilon)$, we plotted $t \log (t/\varepsilon)$ for $d_{\mr{evol}}$ by setting $t=\Delta t$ and $\varepsilon = \norm{U_{\mathrm{ref}}^\dagger - U_{\mathrm{evol}}}_\mr{F}$, as shown in Fig.~\ref{fig: scaling_1column.png}, where the eight-qubit model was chosen. The plots look linear for each $\Delta t$, suggesting a $\log(1/\varepsilon)$ dependence for $\varepsilon$. 

Note that we numerically checked the dependence on $t$ as Ref.~\cite{Karacan2026-zh} using a least squares fitting for
\begin{equation}
    d_{\mr{evol}}(t,\varepsilon) = (u_0+u_1 t^{u_2})\log(\frac{t}{\varepsilon}) + (u_3+u_4 t^{u_5}),
\end{equation}
where $u_0, u_1,\dots, u_5$ are parameters. We obtained the values $u_0 = 1.40, u_1 = 5.50, u_2 = 1.44, u_3 = 6.78, u_4 = 14.95, u_5 = 1.19$ with $R^2 = 0.89$.
\clearpage

\bibliographystyle{apsrev4-1} 

\bibliography{TPET}

\end{document}